# Repeatability of image quality in very low field MRI

Pavan Poojar[1], Kunal Aggarwal[1],
Marina Manso Jimeno[2], and Sairam Geethanath[1,2*]

[1]Accessible Magnetic Resonance Laboratory, Biomedical Imaging and Engineering Institute, Department of Diagnostic, Molecular and Interventional Radiology, Icahn School of Medicine at Mount Sinai, New York, NY, United States

[2]Columbia Magnetic Resonance Research Center, Columbia University, New York, NY, United States

**Correspondence to:**

**Sairam Geethanath, Ph.D.**

Accessible Magnetic Resonance Laboratory, Biomedical Imaging, and Engineering Institute,

Department of Diagnostic, Molecular, and Interventional Radiology

Icahn School of Medicine at Mt. Sinai, New York, NY, United States

E-mail: sairam.geethanath@mssm.edu

Phone: 717-590-0997

# Abstract

**Background**: Low-field magnetic resonance (MR) has emerged as a promising alternative to high-field MRI scanners, offering several advantages. One of the key benefits is that low-field scanners are generally more portable and affordable to purchase and maintain, making them an attractive option for medical facilities looking to reduce costs. Low-field MRI systems also have lower radiofrequency (RF) power deposition, making them safer and less likely to cause tissue heating or other safety concerns. They are also simpler to maintain, as they do not require cooling agents such as liquid helium. However, these portable MR scanners are impacted by temperature, lower magnetic field strength, and inhomogeneity resulting in images with lower signal-to-noise ratio and geometric distortions. It is essential to investigate and tabulate the variations in these parameters to establish bounds so that subsequent in vivo studies and deployment of these portable systems can be well-informed.

**Purpose:** To investigate the repeatability of image quality metrics such as SNR, image uniformity, and geometrical distortion at 0.05T over ten days and three sessions per day

**Methods:** We acquired repeatability data over ten days with three sessions per day. The measurements included temperature, humidity, transmit frequency, off-resonance maps, and 3D turbo spin echo (TSE) images of an in vitro phantom. This resulted in a protocol with nine pulse sequences. We also acquired a 3T data set for reference. The image quality metrics included computing SNR, image non-uniformity, and eccentricity (to assess geometrical distortion) to investigate the repeatability of 0.05T image quality. The image reconstruction included drift correction, k-space filtering, and off-resonance correction. We computed the coefficient of variation (CV) of the experimental parameters and the resulting image quality metrics to assess repeatability.

**Results:** The range of temperature measured during the study was within $1.5^0C$. The off-resonance maps acquired before and after the 3D TSE showed similar hotspots and changed mainly by a global constant. The SNR measurements were highly repeatable across sessions and over the ten days, quantified by a CV of 4.9%. The magnetic field inhomogeneity effects quantified by eccentricity showed a CV of 13.7% but less than 5.1% in two of the three sessions over ten days. The use of conjugate phase reconstruction mitigated geometrical distortion artifacts. The repeatability of image uniformity was moderate at 10.6%, with two of three sessions resulting in a CV of less than 7.8%. Temperature and humidity did not significantly affect SNR and mean frequency drift within the ranges of these environmental factors investigated.

**Conclusions:** We found that humidity and temperature in the range investigated did not impact SNR and frequency. Based on the coefficient of variation values computed session-wise and for the overall study, our findings indicate high repeatability for SNR and magnetic field homogeneity; and moderate repeatability for image uniformity.

## Introduction

Magnetic Resonance Imaging (MRI) is a life-saving technology widely used to investigate the human brain. However, two-thirds of the world's population, particularly those in low-resource settings, lack access to MRI due to the complex requirements of high-field systems, such as infrastructure, engineering, and electrical power [1].

Very low-field MR (<0.1T) has emerged as a promising alternative to high-field MRI scanners, offering several advantages [2–4]. One of the key benefits is that low-field scanners are generally more portable and affordable to purchase and maintain, making them an attractive option for medical facilities looking to reduce costs. Low-field MRI systems also have lower radiofrequency (RF) power deposition, making them safer and less likely to cause tissue heating or other safety concerns [5]. They are also simpler to maintain, as they do not require cooling agents such as liquid helium. Therefore, low-field MRI scanners offer a more portable, accessible, and cost-effective means of accessing vital medical imaging services, particularly in low-resource settings, and have the potential to revolutionize healthcare delivery.

Yip and Konova [6] showed that more frequent and focused neuroimaging scans could allow monitoring brain changes more effectively over time, which can lead to the development of personalized treatment plans for patients. Low-field MRI can address this need for a dense temporal sampling of neuroimaging data. Therefore, continued research and development in this area could significantly impact the accessibility and quality of healthcare globally. However, challenges remain in integrating low-field MRI into clinical practice and ensuring their widespread adoption in both high and low-resource settings[1]. Low-field MRI systems also have a lower signal-to-noise ratio (SNR), which can affect image quality and accuracy. Low spatial resolution is another disadvantage of low-field MRI systems limiting their ability to detect small anatomical details such as blood vessels or small lesions. These limitations result in longer scan times, increasing the risk of patient motion during the scan and reducing image quality. Importantly, low-field MRI systems are susceptible to environmental factors such as temperature, humidity, and electromagnetic interference (EMI) and system factors such as main magnetic field inhomogeneity and gradient heating [3,7–9], which cause image artifacts and degrade image quality. Repeatable and consistent imaging is critical for diagnosing diseases or monitoring therapy progression [9–14]. This is more relevant in the context of these low-field scanners as inconsistencies caused in scanner operation and image quality are due to environmental and system factors changes, especially in permanent magnet constructions. However, the repeatability of low-field MR imaging has not yet been investigated.

In this work, we performed a repeatability study on an in vitro phantom (ProMRI, ProProject, USA) at 0.05T using 3D turbo spin echo (TSE) imaging over ten days and three sessions daily (30 scan sessions). We investigated the precision of very low field MRI (0.05T) measurement quantified by SNR, image uniformity, and geometrical distortion and; investigated the effect of temperature and humidity on image quality. This data would help determine the reliability of the 0.05T scanner and establish bounds of performance that can be used to monitor stable operation. We did not consider the repeatability of contrast (T1w, T2w) as that is more optimally investigated in healthy human volunteers.

## Materials and Methods

**Overview of the study:** We performed the repeatability study using the in-vitro phantom (Pro-MRI by Pro Lab) on a 0.05T scanner (MultiWave Technologies, France). The scanner and hardware specifications are described in detail in Ref. [8] Three scanning sessions were conducted daily: Session 1 began at 11 AM, Session 2 at 2 PM, and Session 3 at 5 PM local time. Each session lasted approximately forty-five minutes. We measured temperature (°C) and humidity (%RH) before and after each Session at three locations. Figure 1 presents a schematic diagram outlining the details of the study, including the Session and time information. Supplementary Figure 1 shows a picture of the 0.05T scanner and the locations where temperature and humidity were measured using a portable digital thermo-hygrometer. The three locations were: in front of the phantom within the bore (location 1), in front of the bore (location 2), and in the center of the room (location 3), shown with orange, yellow, and green circles, respectively. These measurements assessed the potential effect of temperature and humidity on image quality. The device was also compared with a fluoroptic probe (LumaSense, USA) to ensure that the magnetic field did not affect the device's measurements.

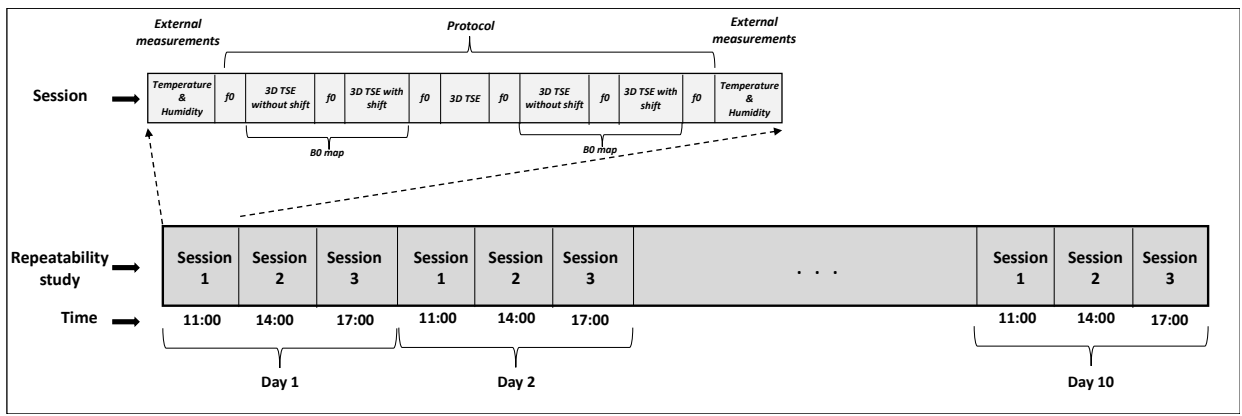

*Figure 1. Repeatability protocol.* *The study was performed over ten days. Three imaging sessions were performed daily, each containing temperature and humidity measurements before and after the scan. The imaging protocol measured transmit frequency, off-resonance mapping, and 3D imaging. The acquisition details are listed in Table 1*

**MR acquisition:** The repeatability protocol utilized in this study included nine sequences. Figure 1(a) shows the protocol details during each Session. The imaging pulse sequences included a 3D turbo spin echo (TSE) and two off-resonance ($\Delta B_0$) maps (before and after 3D TSE). Two 3D TSE sequences with an echo shift of 49µs between them were used to generate the $\Delta B_0$ map, followed by the conjugate phase reconstruction (CPR) method to compensate for off-resonance effects. [15] To find the transmit frequency, a 'findf0' sequence ($f_0$ refers to Larmor frequency, using a pulse and acquire pulse sequence) was employed before and after each sequence, resulting in six such scans. Before beginning the protocol, we measured the noise (no transmit RF pulse, only recording noise) using a 'Monitor noise.' During the repeatability study, we turned off the scanner after each Session (except on day 1) but did not move the scanner and the phantom.

*Table 1. MRI acquisition parameters* at 3T and 0.05T with a spatial resolution of 0.75 x 0.75 x 5 mm$^3$ and 1.5 x 1.5 x 5mm$^3$ respectively.

| MRI acquisition parameters | | |
|---|---|---|
| Field strength | **3T** | **0.05T** |
| Sequence | 2D Turbo spin echo | 3D Turbo-spin-echo |
| TR (ms) | 6000 | 500 |
| TE (ms) | 103 | 20 |
| Echo train length (ETL) | 18 | 4 |
| Bandwidth (kHz) | 56 | 40 |
| Field of view (mm$^3$) | 192 x 192 x 125 | 230 x 230 x 125 |
| Matrix | 256 x 256 x 25 | 155 x 155 x 25 |
| Scan time (min:sec) | 2:02 | 8:01 |

During each session, we measured $f_0$ before and after each scan to evaluate the effect of frequency drift on the image. We scanned the phantom on the 0.05T scanner for ten days and on a Siemens 3T Skyra scanner once. Table 1 lists the acquisition parameters of the protocol at 3T and 0.05T for the 2D multi-slice and the 3D TSE sequences, respectively. The acquisition parameters for the 3D TSE sequence with and without echo shift were the same as the 3D TSE, except for the bandwidth (BW), which was 50 kHz. We used a higher BW for $B_0$ mapping to reduce geometric distortion with shorter readouts at the cost of lower SNR. The total scan time for the protocol was approximately 45 minutes. We acquired 25 slices with a resolution of 5 mm (125mm slab thickness) along the third dimension for a phantom that had a height of 90 mm. Hence, we discarded the first four and last five slices outside the phantom and contained noise.

**Phantom and image quality metrics:** The Pro-MRI phantom solution contained ten mmol of Nickel Chloride and 75 mmol of Sodium Chloride. We evaluated the SNR, image uniformity (IU), and geometric accuracy (GA). These parameters were selected to assess the repeatability of the scans across different imaging sessions over ten days and involved measurements on different slices. Supplementary Figure 2(a-c) displays the representative specific slices, without off-resonance correction, chosen to evaluate SNR, image uniformity, and geometric accuracy in this study. The Pro-MRI phantom's construction is similar to the American College of Radiology phantom, a standard phantom used in MRI. We chose slice 10 for SNR measurement because it lacked apparent fine features or structures. The SNR was computed as the ratio of the mean signal intensity of the phantom ROI to the standard deviation of the background noise. To evaluate IU, we chose slice 16 and computed the standard deviation with the phantom region of interest (ROI, red ROI in Figure 2c; the lesser the standard variation, the more uniform the intensity). For GA, we chose slice number 8 and calculated the eccentricity of the phantom ROI, which measures the degree of deviation from a perfect circle. This calculation was performed before and after applying $B_0$ correction on the 3D TSE data to quantify the benefit of the off-resonance correction.

**Image reconstruction:** Figure 2(a-c) shows the reconstruction, post-processing, and image analysis pipelines. The k-space data underwent preprocessing steps, including drift correction and k-space filtering (squared sine-bell). Drift correction was necessary to correct any signal drift over time, which could lead to image blurring and ghosting, and high-frequency noise was suppressed using k-space filtering. Next, the image was reconstructed using a fast Fourier transform (FFT). However, the off-resonance effects distorted the reconstructed 3D TSE images. $B_0$ correction included the CPR method implemented in our open-source Python toolbox Off-resonance Correction OPen soUrce Software (OCTOPUS) [15]. The results were benchmarked against the results of the code used in ref. [16]. The off-resonance correction process consisted of four steps. In the first step, image reconstruction from two 3D TSE k-space datasets, with a difference in echo time of 49us. We calculated the $B_0$ map from these images, including phase unwrapping, Gaussian smoothing (sigma=21, refer to Supplementary Figure 3) to reduce noise in the corrected images, and background masking (manually) using the magnitude images.

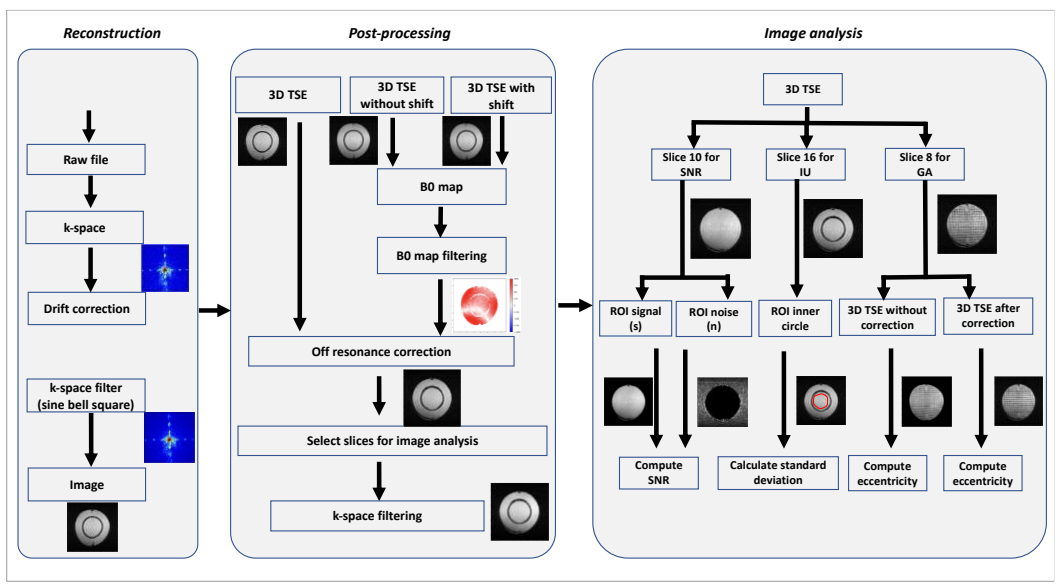

*Figure 2. **Image reconstruction, processing, and analysis pipelines**. (a) The data were put through a series of steps to extract and reorder the k-space from the scanner and corrected for thermal drift and denoising in k-space. (b) The post-processing involved off-resonance correction and low-pass filtering to yield the images for analysis. (c) Different slices were chosen to test for signal-to-noise ratio (SNR), image uniformity, and geometrical distortion and were manually segmented to provide input for computing the relevant image quality metric.*

Finally, we performed the $B_0$ correction using both CPR implementations. We first masked the $B_0$ map to remove the unwanted noise (background). This 3D mask was manually generated from no-shift 3D TSE data based on thresholding. All 30 datasets of the repeatability study underwent this process. We used OCTOPUS for off-resonance correction as it was built in-house and is open source.

**Transmit frequency ($f_0$) and difference in off-resonance measurements: We obtained six f0 values for each session** by measuring the frequency before and after each sequence. We graphed a box plot to compare the frequency drift over ten days and across all sessions. We calculated the mean difference in off-resonance before and after the 3D TSE sequence. We plotted the mean $f_0$ (average of two consecutive $f_0$ values – before and after an imaging pulse sequence), temperature, and humidity for all ten days and the three Sessions.

**Image analysis:** We observed that the slice selected for IU had a ghosting and ringing-like artifact caused due to the stimulated echos and residual magnetization of the long $T_2$ components that we verified by varying the echo train length (ETL; refer to Supplementary Figure 4), which we suppressed by applying k-space filtering. We computed the IU before and after the filtering. We used the Gaussian filter for k-space filtering and optimized the sigma values, varying from 15 to 25. We chose the optimal sigma value by visually inspecting the difference image obtained by subtracting the original image from the filtered image. To evaluate the accuracy of the distortion correction, we calculated the eccentricity of one slice (slice 8) of the phantom before and after correction using the scikit-image library[9]. In the case of no geometric distortion, the eccentricity of a circular slice of the phantom should be equal to 0. We computed the eccentricity of the original image and two correction methods.

**Statistical analysis:** To determine the repeatability of experimental and measured parameters, we computed the Session-wise and overall (all thirty sessions) mean, standard deviation, and coefficient of variation for (i) environmental variables such as the temperature and humidity; MR acquisition parameters such as transmit frequency; and (iii) image quality parameters quantified by SNR, image non-uniformity and eccentricity. A scatter plot was generated to investigate the relationship between (i) temperature and SNR; (ii) humidity and SNR; (iii) temperature and mean frequency; (iv) humidity and mean frequency. The correlation coefficient was computed for each plot to quantify this relationship's strength.

**Results**

***Temperature, humidity, and $f_0$ measurements:*** Figure 3 shows the temperature and humidity measurements. The temperature ranges for all three locations over the thirty sessions were primarily within $1.5^0C$. As expected, the temperature after the scan (dashed) was higher than before (solid), primarily due to gradient heating. The Mean ± SD temperature for all three sessions at the three locations was $23.34 \pm 0.27^0C$, $23.49 \pm 1.24^0C$, and $23.44 \pm 0.29^0C$, respectively. The Mean ± SD of humidity for Locations 1, 2, and 3 were 45.48 ±4.4 %RH, 44 ±4.6 %RH, and 47.6±4.6 %RH respectively. The low values of SD for temperature and humidity suggest minimal variation at each location and across all three locations (10.3% for humidity and 3.2% for temperature, see Supplementary Table 1 for corresponding location-wise numbers). The temperature and humidity measurements recorded a sudden decrease on day seven at all three locations before Session 1, potentially due to the open room door. This allowed the outside temperature and humidity to influence the measurements. However, the door was closed during all the remaining sessions, and the measurements after scanning (see S1A) showed no decrease in measured values. Figure 3(b) shows that Location 2 recorded a higher temperature value on day 3 (see S3A), which could be attributed to an inaccurate reading at Location 2 (edge of the bore), where the fringe magnetic field is the highest. The rest of the experimental conditions were identical to those on other days. Generally, temperatures for the third Session > second Session > first Session are consistent and may be attributed to the room's air conditioning (refer to Figure 4).

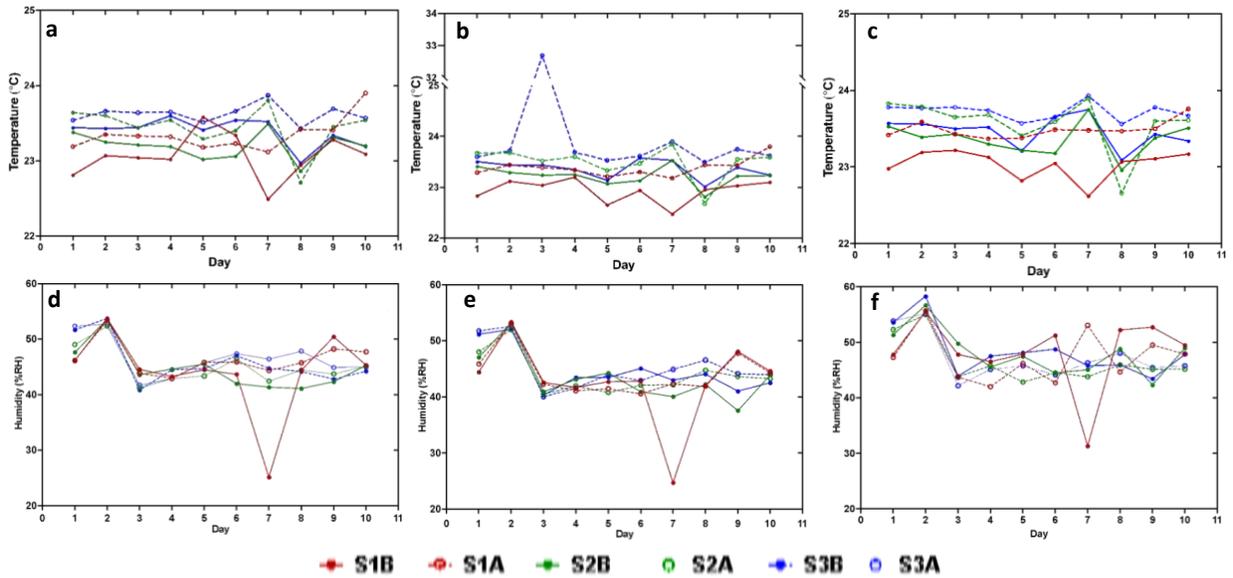

*Figure 3. Temperature and humidity measurements. (a-c), the temperature (in °C) was plotted for ten days at three locations: the center of the bore (a, Location 1), the periphery of the bore (b, Location 2), and the center of the room (c, Location 3). (d-f) shows the humidity for the corresponding three locations. The temperatures and humidity values were measured before the scan (S1B, red filled circle) and after (S1A, red hollow circle). Similarly, the green-filled circles represent Session 2 before and after scanning, and the blue-filled circles with a solid line and the hollow circles with a dashed line represent Session 3 before and after scan measurements.*

Figure 4b shows the box plot of frequency values (MHz) for three sessions over ten days. The plot shows that the frequency was highest during Session 1 and lowest in Session 3, indicating a drift in the central frequency (f0) over time within the same day. This trend is valid for all ten days and validates the temperature trend across sessions seen in Figure 3(a-c). A systematic drift in the central frequency of the MRI scanner can cause distortions and artifacts in the images, especially if the pulse sequence uses the transmit frequency to compute spatial gradient-dependent parameters (slab location, for example) and lead to a loss of image quality. The Mean ± SD of the $f_0$ shift for all three sessions was 28.04 ±14.4 Hz. The high value of the SD compared to the Mean captures the variation in the difference in thermal drift ranges across the three sessions seen in Figure 4.

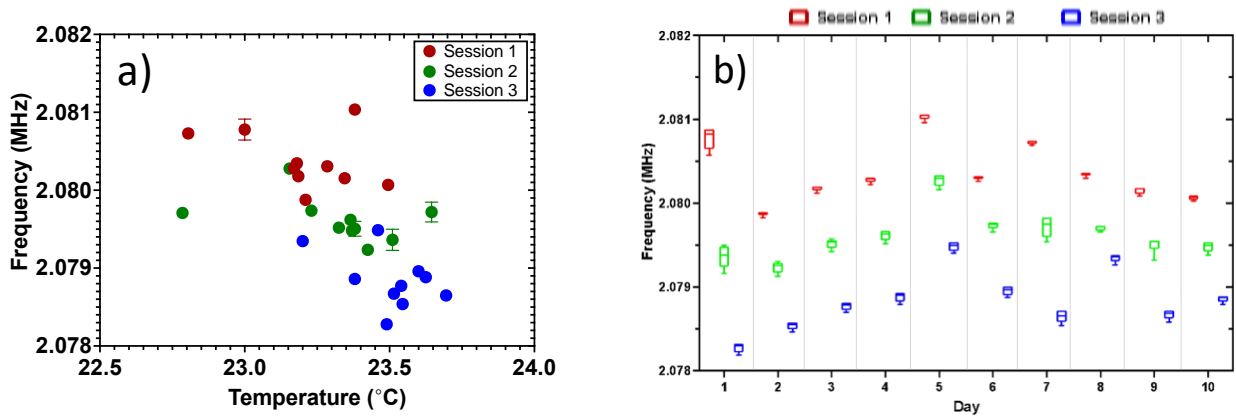

*Figure 4. Temperature dependence of transmit frequency in phantoms.* a) Plot shows that the average transmit frequency per session decreases with an increase in temperature, and session 1 had the least temperature and the highest transmit frequency > session 2 > session 3; b) the box plot shows the range of transmit frequencies for the six measurements obtained in each session.

**$B_0$ mapping:** Figure 5(a,b) shows the off-resonance maps for one slice (slice number 5) for three Sessions over ten days before and after the 3D TSE acquisition. This slice was chosen as it contained a fine structure and inserts with different signal intensities (see Supplementary Figure 2). Figure 5(c) shows the difference in off-resonance maps obtained before and after the 3D TSE acquisition. It can be seen that the maps are in a similar range except for day 1, Session 1 (shown with black arrow), and day 10, Session 1 (shown with green array). The shape of the off-resonance map for day 1, Session 1, is different from the rest of the B0 maps because the scanner was active (turned on) before the start of Session 1 and could be the reason for obtaining a distinct off-resonance compared to the rest of the maps. The off-resonance was saturated with only positive values with the mean value of 19.971 kHz, as shown in Figure 5(b) (day ten, Session 1). One of the possible reasons for this could be electromagnetic interference during the scanning session. The difference in off-resonance values is close to zero, indicating no significant difference between the two measurements, which were acquired before and after the 3D TSE scan. The changes in off-resonance maps were essentially a global shift rather than a hopping of the hotspots in off-resonance.

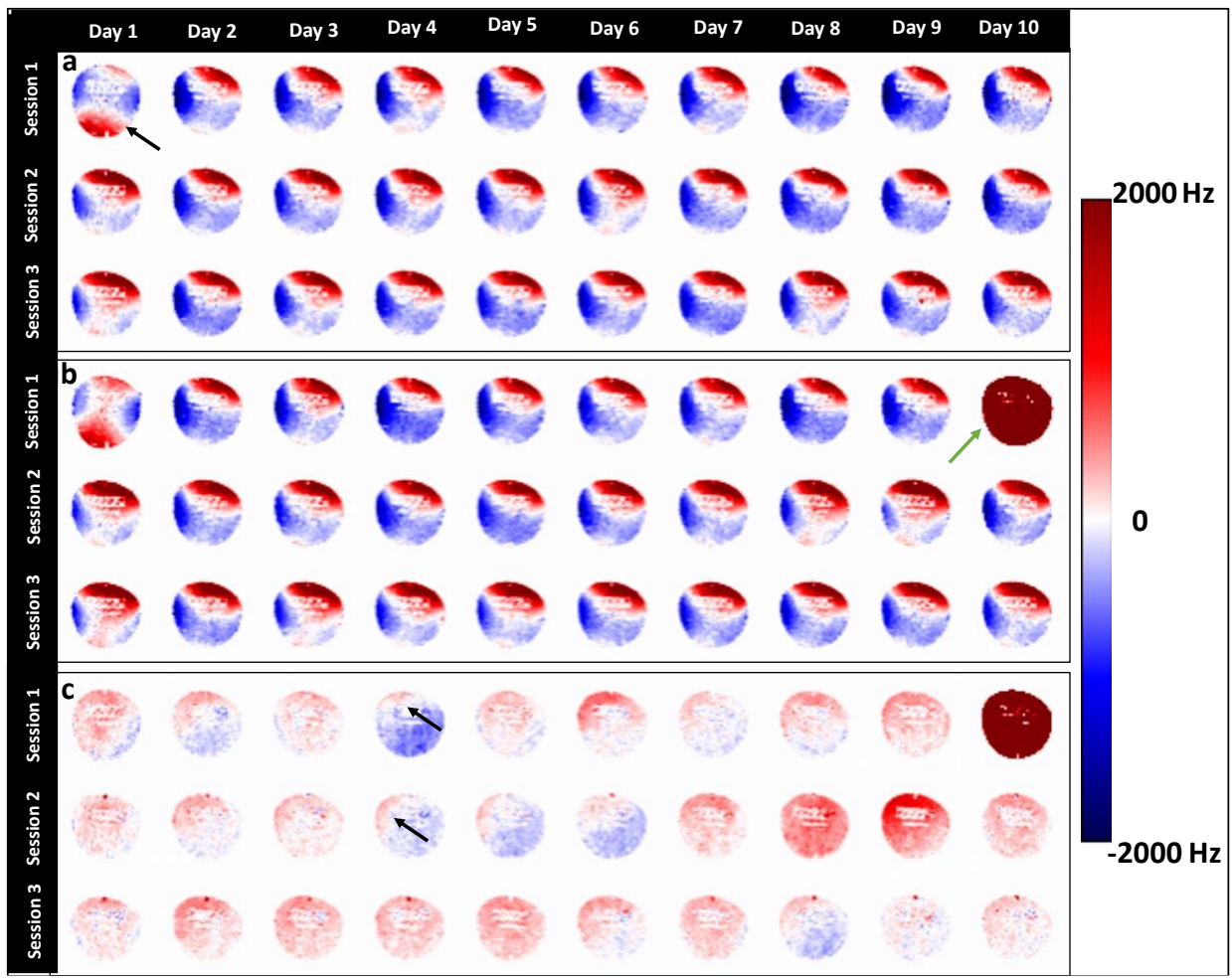

*Figure 5. Consistency of magnetic field homogeneity at 0.05T a) Off-resonance maps for the ten days and three sessions before the 3D turbo spin echo (TSE) scan; b) corresponding off-resonance maps after the 3D TSE scan that do not show any changes in hotspots compared to off-resonance maps in a); c) the difference between a) and b) show examples of changes in off-resonance that do not show a constant difference.*

This can be deduced by the presence of similar red and blue regions in Figure 5(a,b) and the difference depicting significantly either red (positive shift) or blue (negative shift). However, examples like those shown by the black arrows in Figure 5c do not have a constant global shift. These examples do not indicate a shift of the off-resonance hotspots (Figure 5b is similar to 5a in terms of hotspots) but that the difference between the two maps is not a constant. Therefore, acquiring a $B_0$ map post-acquisition may be relevant for downstream off-resonance correction if the imaging sequence lasts beyond several minutes. Exploration of the combination of the two off-resonance maps for optimal CPR correction is beyond the scope of this work. Figure 6(b) shows the plot for the mean difference in off-resonance maps (Hz) over ten days and all three sessions. These values were generally lesser than 400Hz compared to the dynamic range of 4000Hz (~10%).

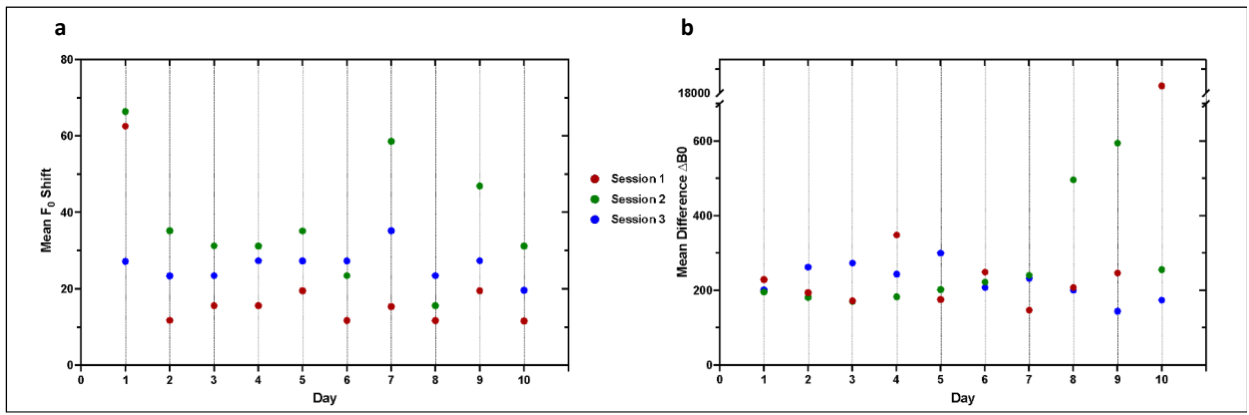

*Figure 6. Intra-session variation in frequency shift and off-resonance a) the mean frequency shift before and after the 3D turbo spin echo (TSE) shows session 1 has the least frequency shift and b) the mean difference in off-resonance before and after 3D TSE does not show a session-wise trend. Day 10 Session 1 shows a significantly higher value, as visualized in Figure 5*

The $B_0$ maps shown here were masked (removed background) and followed by filtering (Gaussian filter). Supplementary Figure 5 (a-c) shows the unfiltered off-resonance maps for one slice over ten days and 3 Sessions. The obtained maps were noisy due to the high BW (50 kHz) acquisition. Supplementary Figure 6(a,b) show the off-resonance maps (16 slices) for unfiltered and filtered 16 slices from day one and Session 1. The first and last slices show high noise as they are away from the isocenter coupled with the low SNR acquisition at 50kHz.

***Off resonance correction:*** All datasets in the repeatability study showed visibly lesser geometric distortion using both CPR implementations by qualitative inspection of the images before and after correction (Supplementary Figure 7). There was no appreciable difference between the two CPR implementations. The eccentricity calculation showed that the phantom's circularity increases after correction, indicating a reduction in geometric distortion. We chose to continue with CPR-OCTOPUS as it is open-source and was home-built.

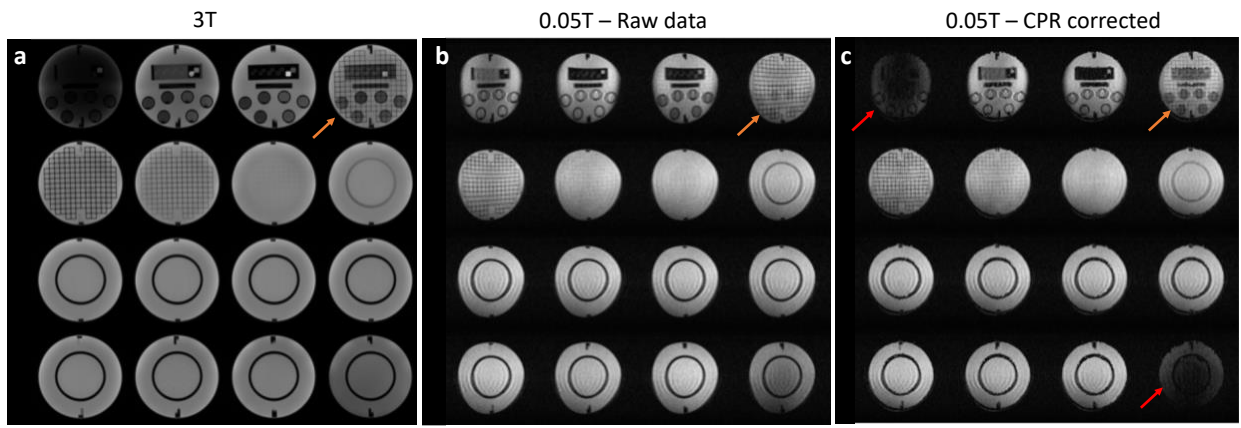

*Figure 7. Image quality comparison a) 3T images from a 2D turbo-spin echo (TSE) sequence of the Pro-MRI phantom at 0.75 x 0.75 x 5mm$^3$ resolution; b) 3D TSE images at 0.05T with no off-resonance correction at 1.5 x 1.5 x 5mm$^3$ resolution, note the offset in the slice direction (orange arrows) and geometrical distortion due to off-resonance; c) off-resonance corrected images reduce geometrical distortion and correct for the slice shift but suffer from a signal loss in the first and last slices (red arrows) due to low signal-to-noise ration off-resonance maps (also see Supplementary Figure 6).*

***Qualitative images:*** Figure 7(a) shows the 2D multi-slice TSE images obtained from Siemens 3T Skyra. These images were considered the gold standard and acquired only once. Figure 7(b,c) shows the 3D TSE images obtained from 0.05T scanner, and off-resonance corrected images using the CPR-OCTOPUS method. The differences in spatial resolution and geometric distortion between the gold standard 3T and 0.05T can be observed by visually comparing Figures 7 (a,c). These image panels show 16 slices from day 1 of Session 1. Slice 4 in all three image panels with an orange arrow indicates partial volume. The first and last slices in corrected images show low SNR (shown in red arrow) in corrected images as well as a shift in the slice direction. This may be attributed to a phase shift applied by the off-resonance correction in the slice direction. The first row images from the 0.05T scanner are more distorted than other slices as these slices are away from the isocenter (see Supplementary Figure 4b).

*Image analysis:* Figure 8(a) shows the SNR over ten days for three Sessions. The SNR is lowest for Session 3 compared to Sessions 1 and 2 in 6 out of 10 days. This is in line with the observation that Session 3 witnessed the highest temperature (Figures 3(a-c)) and lowest Larmor frequency (Figure 4). From Figure 3, it can be observed that the mean temperature increases as the day progress. Therefore, we can deduce that as the temperature increases, SNR decreases. SNR is reduced by thermal noise resulting from the electrical resistance of the imaging coil, which increases with temperature and consequently reduces the SNR. This effect was also reflected in the Mean ± SD of SNR for Sessions 1, 2, and 3 over the ten days, which were 22.4±1.2, 22.05±0.1 and 21.6±1 respectively. However, these similar values (within the range of 1 a.u. of SNR) indicate the repeatability of SNR over the three sessions across ten days. Figure 8(b) shows the plot for IU of the 3D TSE on one slice over ten days and three Sessions. The plot shows that the SD of the intensity values of the ROI were within 0.05 for all ten days. This indicates that the intensity values of the images acquired over ten days using the 0.05T scanner were similar and did not vary significantly, except for day 1. This may be attributed to the scanner not being turned off before operation, as we also see a difference in off-resonance maps (eccentricity) and mean $f_0$ shift for the data from day 1. The IU was computed on the filtered image (Gaussian filter). Supplementary Figure 3 shows the effect of different sigma values on the image. The difference image (last column) was generated by subtracting the filtered image from the original image. As the sigma decreases, the artifact reduces, but the blurring increases. Hence, we chose the sigma value to be 21 based on visually inspecting the different images.

Figure 8 (c-e) shows the eccentricity plot for all three Sessions for the original (without correction, red circle) along with the corrected image using two methods (OCTOPUS - blue circle and CPR - green circle). In all cases, the eccentricity of the corrected image was lesser than the uncorrected, which indicated that the corrected images were closer to a circular shape. The eccentricity decreased by 15.74 ± 3.24% and 16.62 ± 6.99% after the correction (CPR and CPR-OCTOPUS) methods were used, respectively. This discrepancy may be attributed to the code-optimization process during linear algebra operations. However, the ranges of correction significantly overlap.

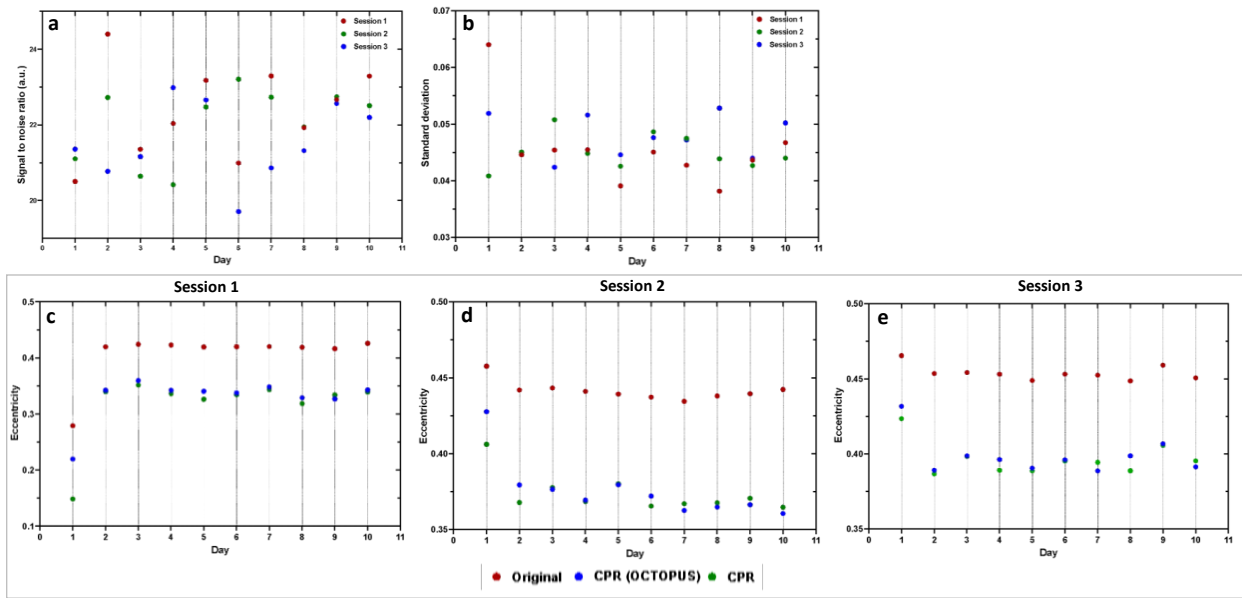

*Figure 8. Image quality repeatability at 0.05T a) signal-to-noise ratio for the three sessions and ten days measured using slice 10; b) image uniformity quantified by the standard deviation (ideal case = 0) in a uniform intensity slice (#16); c-e) geometric distortion quantified by eccentricity (ideal case = 0) for each session over ten days without (original) and with off-resonance correction using two implementations*

***Statistical analysis:*** Table 2 lists the session-wise and the overall mean, standard deviation, and coefficient of variation (CV) for the image quality metrics investigated in this study. The session-wise CV for SNR varied between 4.7% to 5.4%, with a CV of 4.9% across the three sessions over ten days. This indicates that the SNR of the images produced at 0.05T is highly repeatable. The image uniformity quantified by the standard deviation in slice 16 had a wider range of ~8% for session-wise CV, with the overall CV at 10.6%. This moderate repeatability value can be attributed to this metric's fractional dynamic range, significantly impacting percentage changes. Secondly, Gaussian filtering also affects image uniformity, which can be improved using acquisition methods (see Discussion section). The repeatability measures of eccentricity for the uncorrected data were 11%, 1.4%, and 1.1% for the three sessions. The CPR-corrected data had CVs of 18.8%, 5.1%, and 3.2%. The difference in CVs between the first and the other two sessions was due to the significant difference on day 1 when the scanner was not turned off before operation (Figure 8(c), 6(a)). These measurements indicate that the eccentricity values were highly repeatable (highest values of 1.4% and 5.1% for uncorrected and CPR-corrected) if the first session (day 1) was excluded.

*Table 2. 0.05T repeatability statistics* Session-wise and overall mean, standard deviation, and coefficient of variation for the transmit frequency and image quality metrics indicate a variation of less than 1% (for frequency) to 19% for off-resonance-related artifacts.

| 0.05T repeatability statistics | | | | | | | | |
|---|---|---|---|---|---|---|---|---|
| | Session 1 | | Session 2 | | Session 3 | | All sessions | |
| | Mean ± SD | CV (%) | Mean ± SD | CV(%) | Mean ± SD | CV(%) | Mean ± SD | CV(%) |
| Signal-to-noise ratio (a.u.) | 22.36 ± 1.21 | 5.4 | 22.05 ± 0.98 | 4.4 | 21.56 ± 1.02 | 4.7 | 21.99 ± 1.09 | 4.9 |
| Image non-uniformity (a.u.) | 0.045 ± 0.007 | 15.5 | 0.045 ± 0.003 | 6.7 | 0.047 ± 0.004 | 7.8 | 0.046 ± 0.005 | 10.6 |
| Eccentricity raw data (a.u.) | 0.40 ± 0.044 | 11.0 | 0.44 ± 0.006 | 1.4 | 0.45 ± 0.006 | 1.11 | 0.43 ± 0.03 | 7.5 |
| Eccentricity OCTOPUS corrected (a.u.) | 0.32 ± 0.06 | 18.8 | 0.37 ± 0.02 | 5.1 | 0.39 ± 0.012 | 3.2 | 0.36 ± 0.05 | 13.7 |

Figure 9(a) shows the scatter plot for SNR against temperature with a correlation coefficient (r) of 0.48, indicating a weak correlation between temperature and SNR within the investigated range. However, the change in temperature between the Sessions and between the days was within 1.5 ℃, which is very small. This also validates the non-significant difference in Mean SNR compared across the three sessions, which shows a decreasing trend but is within one a.u. change in SNR. Therefore, the SNR at 0.05T will not vary significantly for a temperature range of approximately 1.5$^0$C. Figure 9(b) shows the correlation plot for SNR and humidity. The correlation coefficient is close to zero and positive (r=0.05), indicating a very weak tendency for SNR to change within the range of humidity observed in this study. The correlation is so weak that humidity is unlikely to impact SNR meaningfully. Figure 9(c) shows the mean frequency drift as a function of temperature. The r value of 0.27 indicates a weak positive correlation indicating that the amount of drift is modestly dependent on the absolute temperature value within the ranges investigated. Figure 9(d) shows the correlation plot for humidity versus mean frequency drift with an R-value of 0.05, indicating a weak correlation. These plots suggest that the 0.05T scanner produces images with SNR that are not impacted by temperature or humidity within the ranges observed, enabling repeatability.

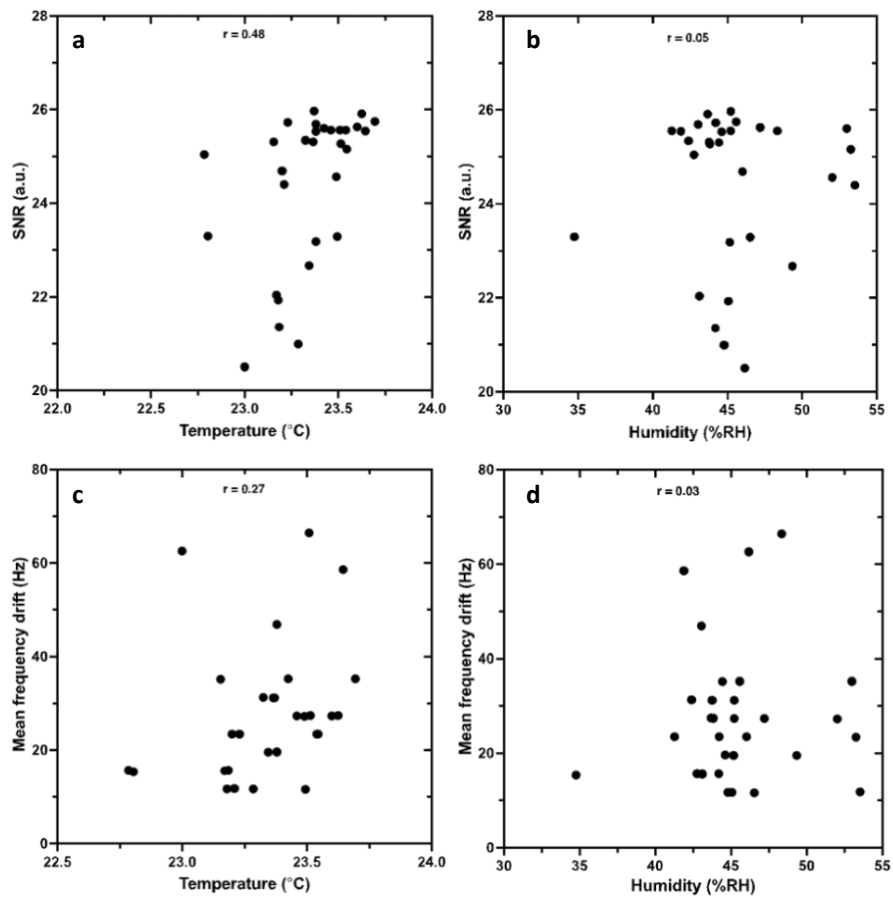

*Figure 9. Image quality stability at 0.05T* The dependence of signal-to-noise ratio on a) temperature and b) humidity; and, correspondingly, the dependence of mean frequency drift on: c) temperature and d) humidity show poor correlations indicating the stability of the 0.05T system, to the ranges of temperature and humidity measured

## Discussion and conclusions

In this study, we evaluated the effect of temperature, humidity, and off-resonance on image quality on an in vitro phantom at 0.05T. The study provided bounds of temperature, humidity, SNR, image non-uniformity, and eccentricity to assess scanner operation and image quality. These bounds can enable better control of the scanner and the resulting images using look-up table methods to deliver highly automated scanner operation [10,17–19]. This repeatability dataset, including k-space and image intensity data, will enable further studies, such as very low-field noise modeling for subsequent denoising algorithms. The frequency–temperature data reproduced similar effects seen in Figure 6 in ref. [8] for phantoms. These bounds provide a benchmark to achieve while performing in vivo acquisitions. The supplementary material also shows repeatability data in the coronal orientation. Supplementary Figure 8 (a, b) shows that the off-resonance was out of range for day six and Session 1, which may have been due to electromagnetic interference. A similar effect was observed for the axial orientation in 10-day Session 1 (see green arrow in Figure 5).

One of the limitations of this study is that we did not scan the phantom near the periphery of the bore, which might be more relevant to brain imaging in such Halbach arrays with smaller bore sizes. However, iso-center imaging relates well to musculoskeletal imaging. We did not also examine image contrast in this study as such an investigation will benefit from imaging healthy volunteers and relevant anatomy. Additionally, we observed ghosting and ringing-like artifacts in some slices caused residual magnetization due to the four ETL of the 3D TSE (Supplementary Figure 4) but did not change our acquisition as we do not expect high proportions of long $T_2$ components in vivo as seen in the phantom. We optimized the sigma value to 21, and the artifact was significantly removed, as seen in Supplementary Figure 3. We did not further investigate k-space filtering as it is ideal for removing it during acquisition. Our sequences already included RF spoiling. This issue can be further addressed by optimizing the TE or applying crushers to reduce the undesired coherences caused by increased ETL. Finally, the total scan time for $B_0$ mapping took 16 minutes per $B_0$ map (2 x 3D TSE scans every 8 minutes), which is not practical for in vivo studies. Therefore, the $B_0$ acquisition must be further time-optimized by acquiring low-resolution off-resonance maps. Our choice was directed by the goal of measuring changes in these maps before and after the image sequence and determining any time-dependent changes in off-resonance hotspots.

In summary, we found that humidity and temperature in the range investigated did not impact SNR and frequency. Our findings indicate a high level of repeatability for SNR and magnetic field homogeneity (quantified by eccentricity, except for day 1, session 1); and moderate repeatability for image uniformity.

## Acknowledgment

The authors would like to thank grant support from the Friedman Brain Institute Research Scholars program at Mt. Sinai, the faculty idea innovation prize at Mt. Sinai, and the Center for Precision Medicine joint collaborative research award – a joint program between Mt. Sinai and Rensselaer Polytechnic Institute (PI: Geethanath).

Supplementary material

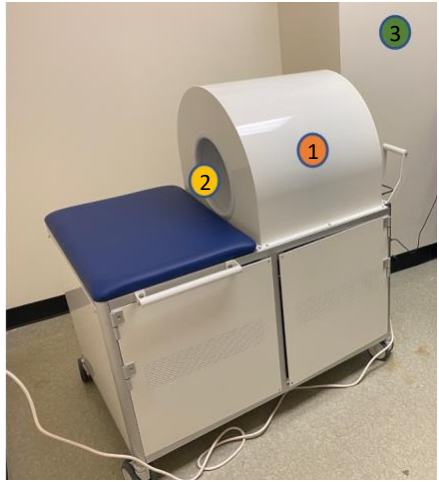

**Supplementary Figure 1**. The portable 0.05T scanner and the three locations used to measure temperature and humidity 1) inside the bore; 2) near the mouth of the bore; 3) in the room

| SNR (slice 10) | Image non-uniformity (slice 16) | Geometric accuracy (slice 8) |

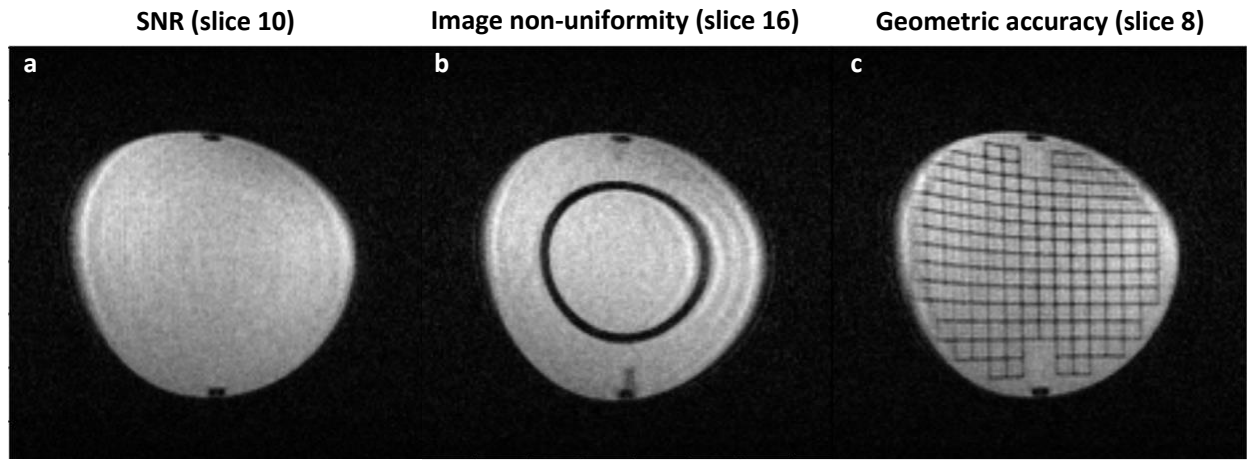

**Supplementary Figure 2.** Raw data (not corrected for off-resonance) slices considered for image quality metrics: a) signal-to-noise ratio – slice 10; b) image non-uniformity quantified by the standard deviation inside the inner circle (ideal = 0)– slice 16; c) geometric accuracy quantified by eccentricity (ideal = 0) – slice 8

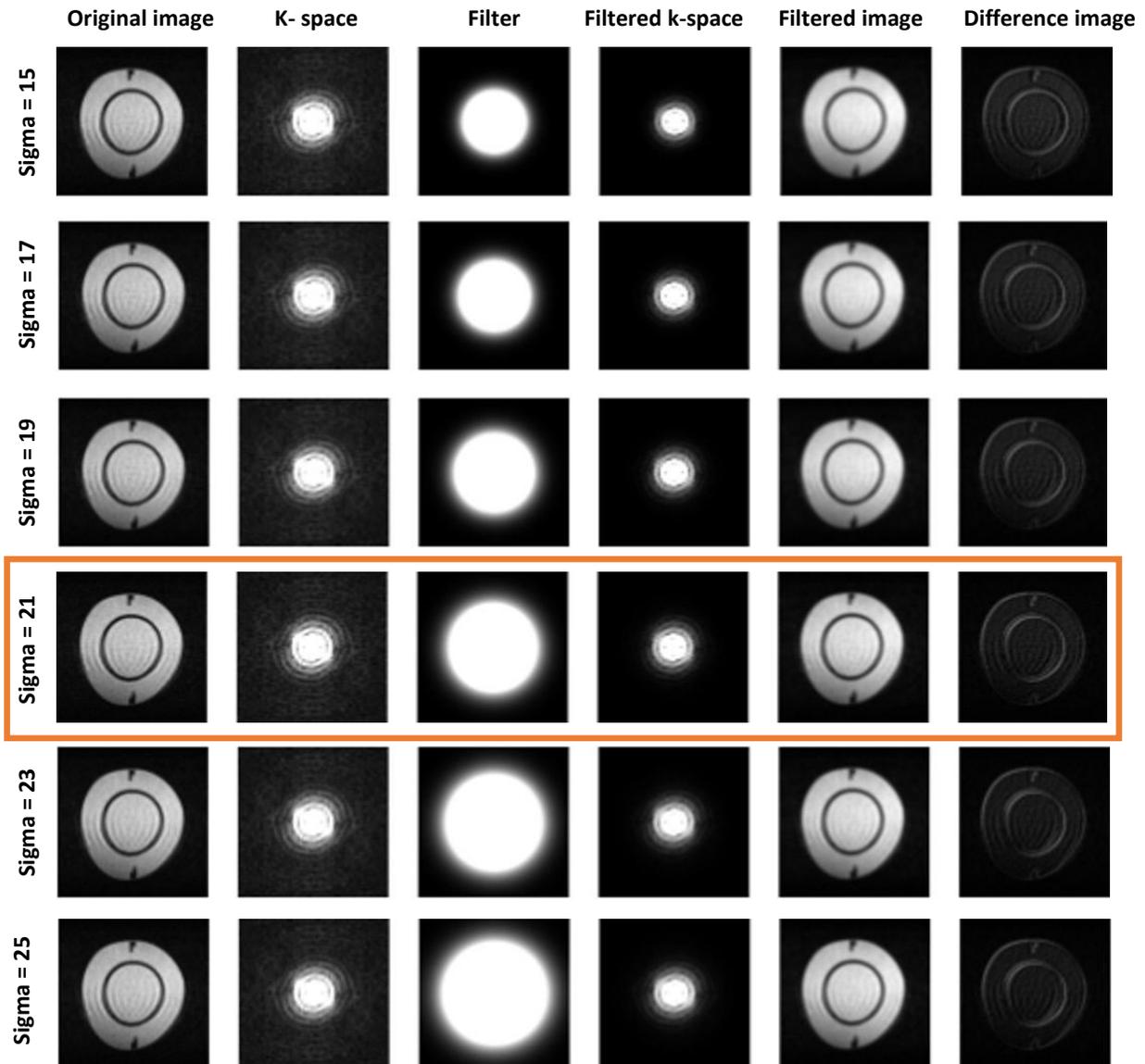

**Supplementary Figure 3.** The choice of the standard deviation for the Gaussian filter was based on the balance between smoothening the ringing artifact and the loss of details, by visual inspection.

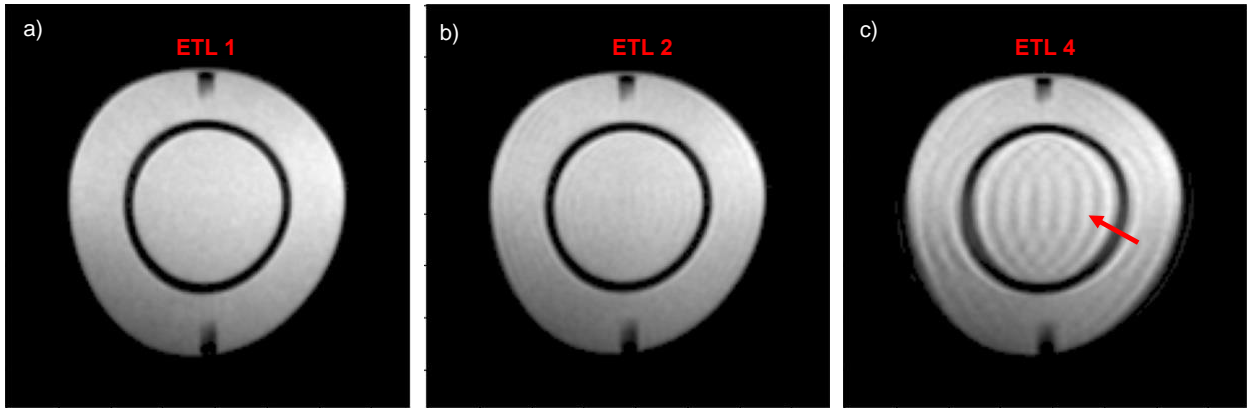

**Supplementary Figure 4.** The effect of echo train length (ETL) on image quality a) ETL =1; b) ETL =2; c) ETL = 4. Increasing ETL increasing the ringing (red arrow) in the image due to lack of suppression of residual magnetizations (stimulated echoes) from previous refocusing pulses.

**0.05T repeatability statistics**

|  | Location 1 | | Location 2 | | Location 3 | | All locations | |
| --- | --- | --- | --- | --- | --- | --- | --- | --- |
|  | Mean ± SD | CV (%) | Mean ± SD | CV(%) | Mean ± SD | CV(%) | Mean ± SD | CV(%) |
| **Temperature (⁰C)** | 23.34 ± 0.27 | 1.2 | 23.49 ± 1.24 | 5.3 | 23.44 ± 0.29 | 1.2 | 23.42 ± 0.75 | 3.2 |
| **Humidity (%RH)** | 45.48 ± 4.35 | 9.5 | 43.99 ± 4.55 | 10.3 | 47.60 ± 4.78 | 9.5 | 45.69 ± 4.70 | 10.29 |

**Supplementary Table 1.** The mean, standard deviation and coefficient of variation in temperature and humidity measurements reported location-wise and for all locations.

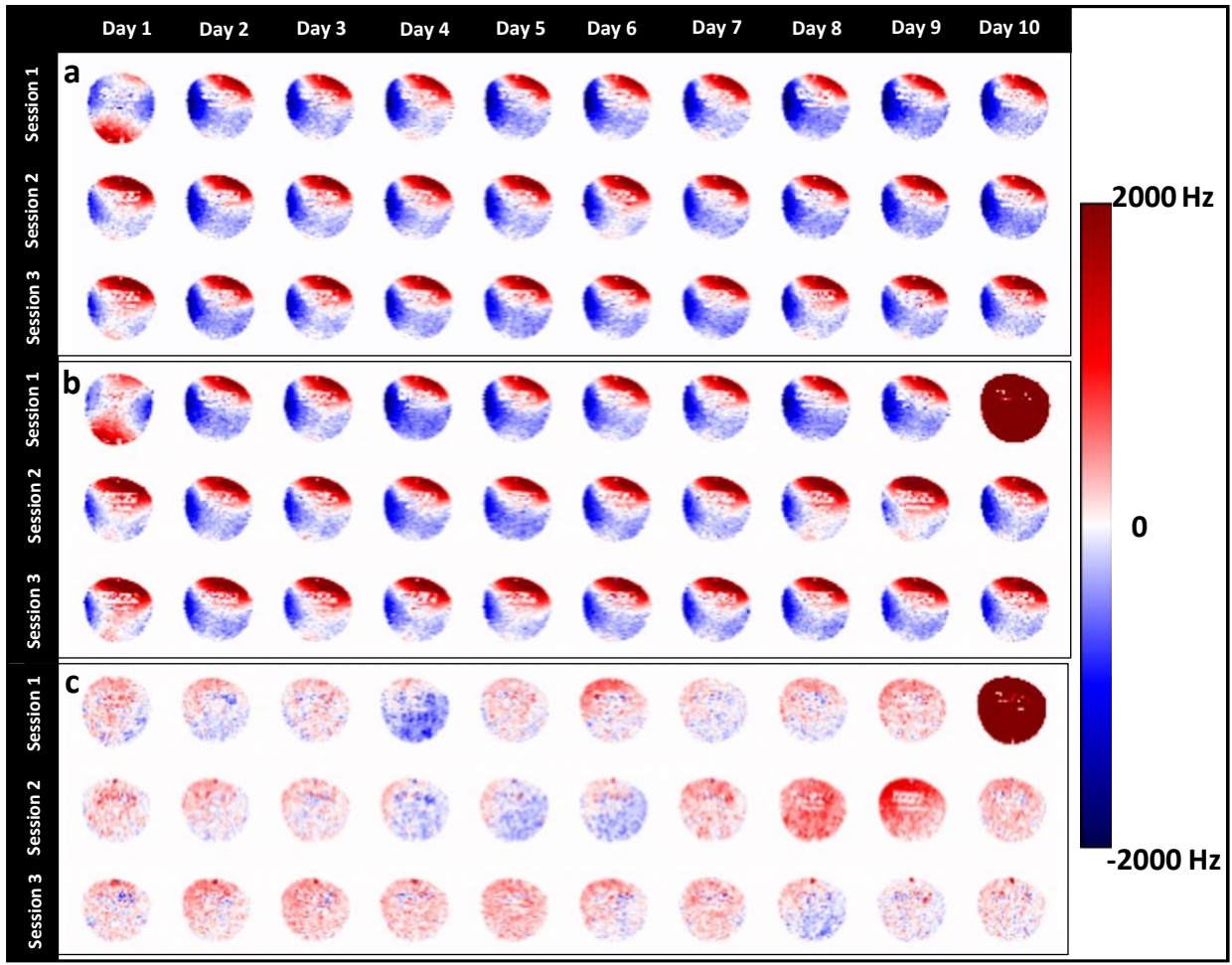

**Supplementary Figure 5.** Unfiltered off-resonance maps for the three sessions and ten days a) before and, b) after the 3D turbo-spin echo sequence; c) difference between a) and b). The noise in these maps is reduced by filtering as shown in Figure 5

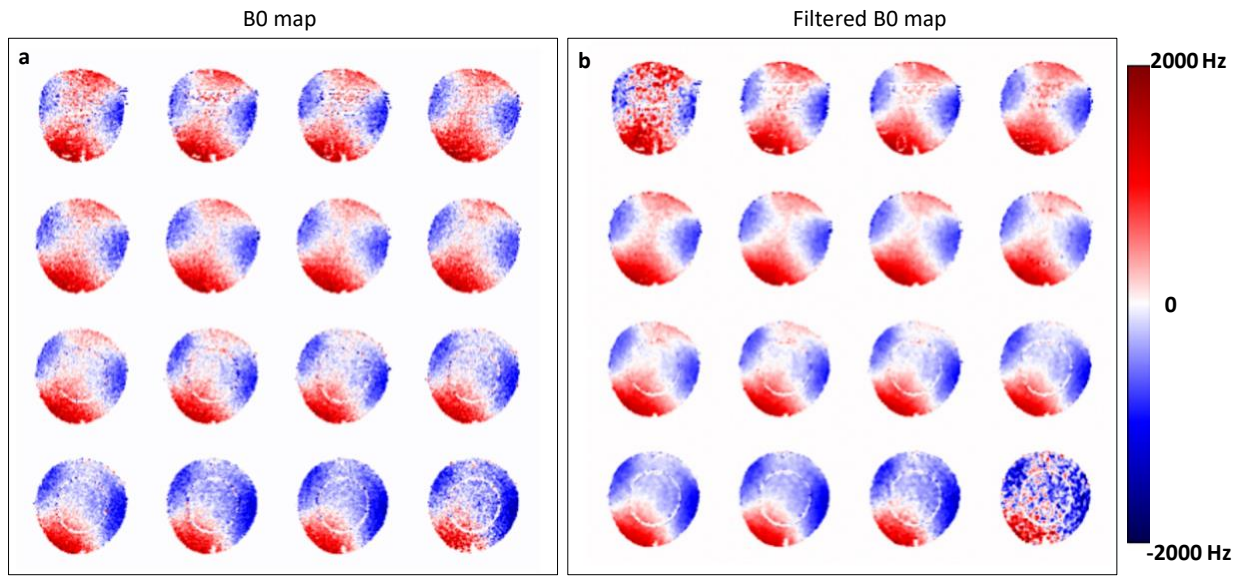

**Supplementary Figure 6.** The effect of filtering on off-resonance maps a) unfiltered and b) filtered off-resonance maps. The low signal-to-noise ratio of the first and last slices can be noted and the corresponding filtered outputs show different off-resonance quality compared to the rest of the slices.

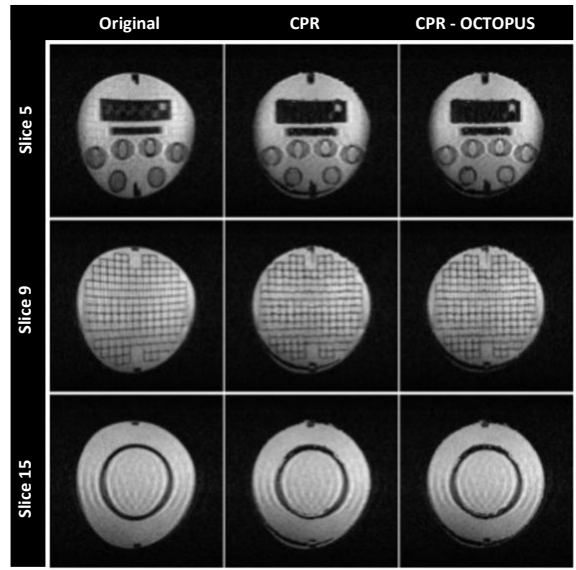

**Supplementary Figure 7.** The comparison of off-resonance correction methods applied on the a) reconstructed image using the available b) conjugate phase reconstruction (CPR) and c) CPR using our custom implementation OCTOOPUS

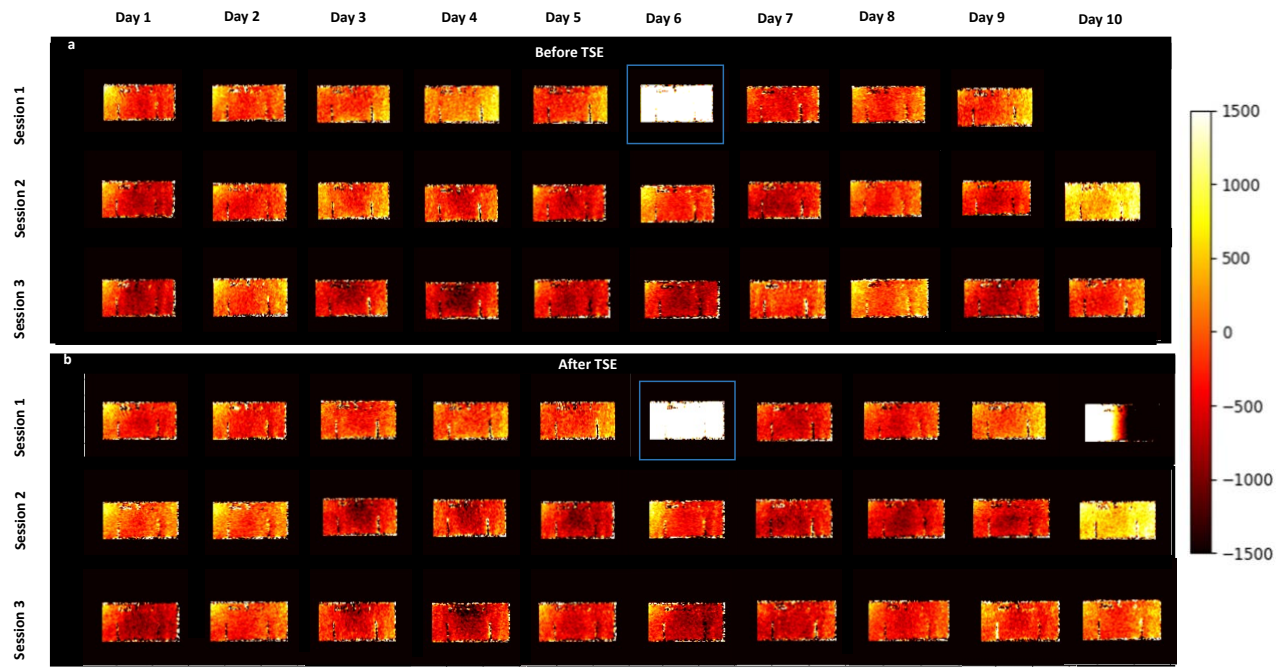

**Supplementary Figure 8.** Off-resonance maps before and after 3D turbo spin echo acquisition acquired in the coronal orientation also shows a single instance of increased off-resonance range similar to the axial measurements in Figure 5.